\documentclass[12pt]{article}
\usepackage{epsf}

\usepackage{epsfig,graphics}
\usepackage {graphicx}
\usepackage {epsfig}
\usepackage{subcaption}
\usepackage {tabularx} 
\usepackage{rotate}	
\usepackage{slashed}
\usepackage{bbm}
\usepackage{color}
\usepackage{tikz}
\usetikzlibrary{decorations.pathmorphing} 
\usepackage{amsmath}
\usepackage{amsfonts}
\usepackage{amssymb}
\usepackage{graphicx}
\usepackage{cite}

\usepackage{fancyhdr}
\usepackage{hyperref}
\usepackage{diagbox}

\newcommand{\bmat}{\left(\begin{array}}
\newcommand{\emat}{\end{array}\right)}

\def\yzero{\smash{\hbox{$y\kern-4pt\raise1pt\hbox{${}^\circ$}$}}}

\def\beq{\begin{equation}}
\def\eeq{\end{equation}}
\def\beqa{\begin{eqnarray}}
\def\eeqa{\end{eqnarray}}

\def\-{\hphantom{-}}

\def\s2{\frac{1}{\sqrt2}}

\def\Dsl{\,\raise.15ex\hbox{/}\mkern-13.5mu D} 
\def\IZ{Z\kern-.4em  Z}

\def\be{\begin{equation}}
\def\ee{\end{equation}}
\def\bea{\begin{eqnarray}}
\def\eea{\end{eqnarray}}
\def\bes{\begin{subequations}}
\def\ees{\end{subequations}}

\catcode`\@=11   
\newdimen\@rotdimen
\newbox\@rotbox  

\def\@vspec#1{\special{ps:#1}}
\def\@rotstart#1{\@vspec{gsave currentpoint currentpoint translate
   #1 neg exch neg exch translate}}
\def\@rotfinish{\@vspec{currentpoint grestore moveto}}
\def\@rotr#1{\@rotdimen=\ht#1\advance\@rotdimen by\dp#1%
   \hbox to\@rotdimen{\hskip\ht#1\vbox to\wd#1{\@rotstart{90 rotate}%
   \box#1\vss}\hss}\@rotfinish}
\def\@rotl#1{\@rotdimen=\ht#1\advance\@rotdimen by\dp#1%
   \hbox to\@rotdimen{\vbox to\wd#1{\vskip\wd#1\@rotstart{270 rotate}%
   \box#1\vss}\hss}\@rotfinish}%
\def\@rotu#1{\@rotdimen=\ht#1\advance\@rotdimen by\dp#1%
   \hbox to\wd#1{\hskip\wd#1\vbox to\@rotdimen{\vskip\@rotdimen
   \@rotstart{-1 dup scale}\box#1\vss}\hss}\@rotfinish}%
\def\@rotf#1{\hbox to\wd#1{\hskip\wd#1\@rotstart{-1 1 scale}%
   \box#1\hss}\@rotfinish}%
\def\rotate{\@ifnextchar[{\@rotate}{\@rotate[l]}}
\def\@rotate[#1]#2{\setbox\@rotbox=\hbox{#2}\@nameuse{@rot#1}\@rotbox}

\catcode`\@=12

\topmargin
-1.5cm
\textwidth
15.5cm
\textheight
23.5cm
\oddsidemargin
0.7cm
\evensidemargin
0.7cm

\begin{document}

\makeatletter
\@addtoreset{equation}{section}
\makeatother
\renewcommand{\theequation}{\thesection.\arabic{equation}}
\pagestyle{empty}
\vspace{-0.2cm}
\rightline{ IFT-UAM/CSIC-24-105}
\vspace{0.5cm}
\begin{center}



\LARGE{Modular  Invariant Starobinsky  Inflation \\
and the Species Scale}
\\[4mm]
\large{Gonzalo F.~Casas,$^\diamondsuit$  Luis E.~Ib\'a\~nez$^{\clubsuit  \diamondsuit}$}
\\[6mm]
\small{$^\clubsuit$  Departamento de F\'{\i}sica Te\'orica \\ Universidad Aut\'onoma de Madrid,
Cantoblanco, 28049 Madrid, Spain}  \\[5pt]
$^\diamondsuit$  {Instituto de F\'{\i}sica Te\'orica UAM-CSIC, c/ Nicolas Cabrera 13-15, 28049 Madrid, Spain} 
\\[3mm]
\small{\bf Abstract} \\ [3mm]
\end{center}
\begin{center}
\begin{minipage}[h]{15.22cm}
Potentials in cosmological inflation often involve scalars with trans-Planckian ranges. As a result, towers of states become massless and their presence pushes the fundamental scale not to coincide with $M_{\rm P}$ but rather with the \textit{species scale}, $\Lambda$. This scale transforms as an automorphic form of the theory's duality symmetries. We propose that the inflaton potential should be 1) an automorphic invariant form, non-singular over all moduli space, 2) depending only on $\Lambda$ and its field derivatives, and 3) approaching constant values in the large moduli to ensure a long period of inflation. These conditions lead to the proposal $V \sim \lambda(\phi, \phi^*)$, with $\lambda = G^{i\bar{j}} (\partial_i \Lambda)(\partial_{\bar{j}} \Lambda) / \Lambda^2$, determining the 'species scale convex hull'. For a single elliptic complex modulus with $SL(2, Z)$ symmetry, this results in an inflaton potential $V \simeq (\text{Im} \tau)^2 |\tilde{G}_2|^2 / N^2$, with $N \simeq -\log(\text{Im} \tau |\eta(\tau)|^4)$, where $\eta$ is the Dedekind function and $\tilde{G}_2$ the Eisenstein modular form of weight 2. Surprisingly, 
this potential at large modulus resembles that of the Starobinsky model. We compute inflation parameters yielding results similar to Starobinsky's, but extended to modular invariant expressions. Interestingly, the number of e-folds is proportional to the number of species in the tower, $N_e \simeq N$, and $\epsilon \simeq \Lambda^4$ at large moduli, suggesting  
that the tower of states plays an important  role in the inflation process.

\vspace{3em}

\end{minipage}
\end{center}
\newpage
\setcounter{page}{1}
\pagestyle{plain}
\renewcommand{\thefootnote}{\arabic{footnote}}
\setcounter{footnote}{0}



\tableofcontents

 %

 \section{Introduction} 
  
 Inflation is a paradigm that has shown to be quite successful in reproducing both qualitative and quantitative 
 facts in modern cosmology (see e.g. \cite{Kinney:2009vz,Baumann:2009ds} and references therein). The simplest inflation models are based on the existence of one (or possibly several) scalar inflaton
 with a potential that has the appropriate conditions to generate a sufficiently long period of highly accelerated expansion.	
Perhaps  the most successful type of potential is that of the Starobinsky model \cite{Starobinsky:1980te}
(see e.g. \cite{DiMarco:2024yzn,Mohanty:2020pfa,Baumann:2014nda} ) which starts with an action of the form
\be
S \ =\ \ \frac {M_{\rm P}^2}{2}\int d^4x \sqrt{-g} \left(R \ +\ \frac {1}{6M^2} R^2\right) \ ,
\label{Staraction}
\ee
where $M$ is some fixed parameter mass scale, numerically of order $M\sim 10^{14}$ GeV, in order to adjust for the 
amplitude of scalar perturbations. After certain Weyl and field redefinitions, the system can be shown to be equivalent to
Einstein's Gravity with a scalar degree of freedom $\phi$ and a scalar potential of the form
	\beq
	V_{Star}\ = \  \frac {3M^2}{8}M_{\rm P}^2  \left| 1\ -\ e^{-\sqrt{2/3}\phi/M_{\rm P}}\right|^2 \ ,
	\eeq
	where $\phi$ is a canonically normalised real scalar.  At the plateau of this potential 50-60 e-folds are obtained starting from $\phi=5-6\ M_{\rm P}$. The results for the inflation parameters may be easily expressed in terms of the number of e-folds and are in
 extremely good agreement with data. 
 This example will turn out to be relevant in our analysis so 
	we give a few more details about it in Appendix \ref{onedimension}.
	
	On the other hand in recent years it has become clear that Quantum Gravity may impose strong constraints on these potentials with large 
	trans-Planckian excursions. 
	Thus we have learnt that when a scalar field takes trans-Planckian values, a tower of states (either KK-like or stringy
 \cite{Lee:2019wij}) should become 
	light. This has been named 'Swampland Distance Conjecture' (SDC) \cite{Ooguri:2006in}
 and has been tested to be true in a large number of String examples, see \cite{Brennan:2017rbf,Palti:2019pca,vanBeest:2021lhn,Grana:2021zvf}
 for reviews on the Swampland Program.
	Furthermore, we have also understood that the fundamental scale of the theory is not the Planck scale but rather the {\it Species Scale}
	$\Lambda$ which can be much lower depending on the number of $N$ states below $\Lambda$ \cite{Dvali:2007hz,Dvali:2007wp,Dvali:2008ec}. 
	The Species Scale in $d$-dimensions  is related to the number of species in a tower $N$ by
	\beq
	\Lambda^{d-2} \ \simeq \ \frac { M_{\rm P}^{d-2}}{N} \ .
	\eeq
	In a given string background, there may be several towers in which case the lowest $\Lambda$ is the actual Species Scale
	(see e.g.\cite{Castellano:2021mmx,Castellano:2022bvr} for casuistics). 
	There is not a unique constant value for the fundamental QG scale  $\Lambda$ which is valid in any direction in the moduli space of the theory. 
	Rather, the species scale is 
	moduli dependent so the value of the fundamental scale of physics varies depending on the moduli direction. 
	In fact it has been recently realised that the species scale is an invariant automorphic form with respect to the duality symmetries
	of the theory, see e.g.  \cite{vandeHeisteeg:2022btw,vandeHeisteeg:2023ubh,Castellano:2023aum,vandeHeisteeg:2023dlw}.
 In fact ${\cal R}^2$ actions with a constant coefficient as in the Starobinsky model are not expected to arise in
 a theory of Quantum Gravity. Rather, the coefficient should be field dependent and an automorphic form in terms of
 the duality symmetries of the theory. See \cite{Lust:2023zql,Brinkmann:2023eph} for recent papers addressing the 
 Starobinsky model from the Swampland/String Theory points of view.

	Given these recent developments, it is interesting to study whether there are simple candidate inflaton potentials which 
	are consistent with these Quantum Gravity conditions, in particular, if we identify the inflaton with some string theory modulus.
	We have looked for a potential in which the vev of the inflaton candidate is trans-Planckian without the associated tower of states ruining the EFT.
	Also, it should take into account that there is a QG cut-off $\Lambda$, which will depend in general on this modulus, and should not be too low
	so that we can still use an EFT approach.  Since we will assume that the inflaton is a modulus, we will assume our scalar potential
	to be modular invariant. All these conditions are relatively general, but there is an additional condition which 
	will narrow quite a lot the search. We will ask for a potential with a plateau structure so that $V\longrightarrow     const$ for large modulus (inflaton)
	to allow a long period of inflation, namely $N_e\sim 50-60$.

	To address this search we first review a number of properties concerning the moduli dependence of the Species Scale 
	in general string vacua which have been recently realized.  One of them is that one can define a lattice of vectors in moduli
	space defined in terms of gradients $\nabla^i\Lambda /\Lambda$ 
\cite{Calderon-Infante:2020dhm,Calderon-Infante:2023ler}.
 One can then define a gradient quantity 
	$\lambda(\phi)\equiv G^{ij}(\nabla_i\Lambda \nabla_j\Lambda)/\Lambda^2$, which has been conjectured (and tested in many string examples  \cite{vandeHeisteeg:2022btw,vandeHeisteeg:2023ubh,Castellano:2023aum,vandeHeisteeg:2023dlw}) 
	to asymptote to non-vanishing constants in the different limits in moduli space.  This function, the 'generator of the scalar convex hull'  is in
	general a modular form of all the moduli in the theory that converges to those constants in the different limits. 
	Taking profit of this general large modulus constant behaviour, we propose to consider 
 a single complex modulus $\tau$ and take  as  the inflaton potential 
 \be
 V(\tau ,\tau ^*) \sim \lambda(\tau,\tau^*)\ .\label{eq:Vlambda}
 \ee
	In principle taking \eqref{eq:Vlambda} looks like a bold assumption, 
 but the consequences of such a choice turn out to be 
	quite surprising and interesting. Furthermore the potential 
 is positive definite and we argue that in the context of supersymmetric theories
 could arise from a D-term potential.
 Taking as an example a $ {\cal N}=2 $ ,  $K_3\times T^2$, Type IIA string compactification as a proxy to evaluate 
the structure of a modular invariant species scale,  one finds the simple expression for the potential,
\be
V \ =\ K^2 |\tau_2{\tilde G}_2|^2\Lambda^4 \ ,
\ee
with $K$ a modulus independent constant, and ${\tilde G}_2$ the weight-2 Eisenstein non-holomorphic form.
As we will show,  at large modulus $\tau_2$ one gets for the (canonically normalized $\phi$ field)
\be
V\ \sim \ |{\tilde G}_2|^2\ \sim | \frac {\pi}{3 } \ -\ e^{-a\phi}|^2   \ .
\ee
Remarkably, this structure is quite similar to that of the Starobinsky model (for $a=\sqrt{2/3}$). Due to this fact, one obtains quite successful results for the 
inflation cosmological parameters like $n_s$ and $r$. However, those parameters are now modular invariant forms, and have an interesting
physical interpretation in terms of the species scale  $\Lambda$ and the number of species $N$.  In particular, at large modulus 
 one obtains that  (in Planck units)
\be
 N_e \ =\ \frac {N}{24a^2} \ , \ 
 \epsilon\ \simeq \  288 a^2\Lambda^4 \ ,\ \eta\ \simeq \ - \ 24a^2 \Lambda^2 \ .
 \ee
So the number of e-folds is determined by the number of states in the tower $60$ e-folds before the end of inflation, and
$\epsilon$ and $\eta$ are given by the species scale, which are then  naturally exponentially small while inflation takes place but become 
larger by the end of inflation.  Thus, unlike in standard inflaton settings, at the initial stages of inflation there was a huge
amount of entropy, associated with the tower of states which are necessarily present, whenever the 
inflaton becomes trans-Planckian.
The entropy stored originally in the tower is eventually transformed into that at the end of inflation.

Let us emphasize that we do not obtain the Starobinsky model in any limit. 
It is just that 
we recover a potential at large modulus quite similar to that model. In particular our inflaton
is the imaginary part of a complex modulus and it does not arise as a dynamical  effect from an ${\cal R}^2$ 
term in the Lagrangian.
Numerically, the results for the inflation parameters at large modulus are identical to those of Starobinsky for $a=\sqrt{2/3}$. 
However as we approach the minimum of the potential at the self-dual point the potential changes very much since, in particular, there
is no wall but a new dual phase opens. So questions like reheating may be in principle quite different.
On the other hand, the full potential is now modular invariant and, unlike the Starobinsky model,
due to this property, the structure is stable under higher dimensional operator corrections. Rather,  this potential is expected to
encapsulate all higher dimensional corrections (as the modular invariant Species Scale functions so far computed do 
 \cite{Green:2010kv,Green:2010wi,Green:1999pv,Pioline:1998mn,Green:1999pu,Green:2005ba}). 
  
  The structure of this paper is as follows. In section \ref{sec: species} we review some recent results about the
  Species Scale as a field dependent automorphic form. In section \ref{sec:proposal} we make the proposal for the 
  inflaton potential being proportional to the generator of the Species Scale gradients convex-hull.
  In section \ref{example} we make the proposal more concrete and compute the proposed inflaton potential as
  a $SL(2,{\bf Z})$ modular invariant form. In section \ref{sec:inflationpara} we compute the inflaton parameters.
  We leave section \ref{conclusions} for some final comments and conclusions. Appendix \ref{onedimension} reviews Starobinsky's model and Appendix \ref{app:modular} collects some useful formulae about $SL(2,{\bf Z})$
  modular forms.

	\section{The species scale and modular symmetries}\label{sec: species}
	
\begin{figure}[h!]
    \centering
    \vspace{1em}
    \includegraphics[width=1\linewidth]{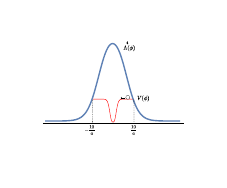}
    \vspace{-7em}
    \caption{Qualitative structure of the potential (in red) in a real slice ($\tau_1=0$). The duality is reflected in the symmetry
    $\phi \leftrightarrow -\phi$. The inflaton needs to start from $\phi_i\gtrsim 5$ so that enough inflation is produced. 
    The bell-shaped line corresponds to the bound $V\lesssim \Lambda^4$ required for an EFT to be consistent. That confines the
    region of the inflaton to a range $|\phi|\lesssim 10/a$. }
    \label{fig:speciesvspotential}
\end{figure}

	In physics configurations in which a large number of species arise, like e.g. the states in a KK tower or the states in a string, 
	the species scale plays an important role. 
	Since different moduli directions vary the structure and the very existence of the towers, it is obvious that the scale of the species should be moduli dependent. It has been emphasized recently 
	\cite{vandeHeisteeg:2023ubh,Castellano:2023aum,vandeHeisteeg:2023dlw} that the 
	species scale should be   (generalized) modular invariant functions of the moduli of the theory. Very explicit results have been obtained for
	maximal supersymmetric theories, which may be obtained by dimensional reduction of M-theory. The idea is to realise that there are higher 
	dimensional operators involving derivative of curvature tensors which are BPS operators
 \cite{Green:2010kv,Green:2010wi,Green:1999pv,Pioline:1998mn,Green:1999pu,Green:2005ba}.
 They appear in loops or at tree level and should be 
	suppressed by the fundamental scale of gravity, thus by powers of the species scale $\Lambda$
 \cite{vandeHeisteeg:2023ubh}
	\beq
	S_{EFT,D} \ =\ \int d^Dx \frac {1}{2\kappa_D^2} \left({\cal R} \ +\ 
	\sum_n\frac {{\cal O}_n({\cal R}) } {\Lambda^{n-2}} \right)\ +... \ ,
	\eeq 
	where ${\cal O}_n({\cal R})$ are higher dimensional operators of mass dimension $n$.
	Thus this provides us with an alternative definition of
	$\Lambda$.  The coefficients of these operators appearing in loops had  been calculated in the past for  a good number of cases
  \cite{Green:2010kv,Green:2010wi,Green:1999pv,Pioline:1998mn,Green:1999pu,Green:2005ba}, so providing us with 
	explicit expressions for the species scale in some instances. For example, in the case of  10D IIB  strings, there is the $SL(2,{\bf Z})$ symmetry
	of the axi-dilaton and there is a higher dimensional term
	\beq
	S_{IIB} \ = \  \frac {1}{\kappa_{10}^2} \int d^{10}x \sqrt{-g} E_{3/2}^{sl2}(\tau) t_8t_8{\cal R}^4 \ ,
	\eeq
	where $t_8$ is some standard  tensor  and $E_{3/2}^{sl2}(\tau)$ is the order-3/2 non-holomorphic Eisenstein series 
 \footnote{ This should not be confused with the holomorphic Eisenstein $SL(2,{\bf Z})$ modular forms described in Appendix \ref{app:modular}.} of
	$SL(2,{\bf Z})$ which is an invariant automorphic form (see e.g. appendices in \cite{Castellano:2023aum} and references therein).
	So one can write for the species scale $\Lambda_{IIB}^6=1/E_{3/2}^{sl2}$ 
 \cite{vandeHeisteeg:2023ubh,Castellano:2023aum,vandeHeisteeg:2023dlw}. 
 The duality symmetries get larger as we compactify on circles so
	that the number of moduli increases and the situation gets more involved. Another simple example is that of maximal supersymmetry in $D=8$
	(or M-theory on $T^3$). In this case the discrete duality symmetry is $SL(2,{\bf Z})\times SL(3,{\bf Z})$ 
 and one has  \cite{Green:2010kv,Green:2010wi,Green:1999pv}
	$\Lambda_{8D}^6 = 1/({\hat E}_{3/2}^{sl3}+2{\hat E}_1^{sl2})$, where ${\hat E}_{3/2}^{sl3}$ and ${\hat E}_1^{sl2}$ are respectively
	the ('regularized') non-holomorphic Eisenstein series of order 3/2 and 1 of $SL(3,{\bf Z})\times SL(2,{\bf Z})$. One can write in particular for the latter
	\beq 
	{\hat E}_1^{sl2} \ =\ 	-\pi \ \log({\rm Im}\tau |\eta(\tau)|^4) \ .
	\eeq
 This automorphic form will have an important place in our discussion.
Indeed, it appears naturally in the case of ${\cal N}=2$  Type IIA CY compactifications, and 
its role has also been explored in detail in this context.
Considering the one-loop ${\cal R}^2$ operator coefficient,
\beq
	S_{EFT,4} \ =\ \int d^4x \frac {1}{2\kappa_4^2} \left({\cal R} \ +\ 
	\sum_n\frac  {{\cal R}^2 } {\Lambda^{2}} \right)\ +... \  , 
 \label{desarrollo}
	\eeq 
in \cite{vandeHeisteeg:2022btw} it was argued that the species scale in Type IIA Calabi-Yau compactifications is related to the 
one-loop topologically free energy $F_1$. A general expression of $F_1$ for the vector multiplet sector reads as follows (up to an additive constant)
\be
F_1 \ = \ \frac {1}{2}(3+h^{1,1}+ \frac {\chi}{12}) K\ - \ \frac {1}{2}\log {\rm det}G_{i{\bar j}} \ .
\label{F1}
\ee
Here $h_{1,1}$ is the number of vector multiples and $\chi$ the Euler characteristic of the CY manifold. 
$K$  and $G_{i{\bar j}}$ are the K\"ahler potential and the metric of the vector multiplet moduli space. 
Thus for this class of theories one can write $N = F_1 +N_0$, with $N_0$ an additive constant.
Consider for example the Enriques CY $K3\times T^2/{\bf Z}_2$.
	It was found in    \cite{vandeHeisteeg:2023dlw}
	\be
	N \ \simeq \ -6\log\left[2\tau_2|\eta(\tau)|^4\right] \ +\  {\tilde N}_0 \ \ ;\ \ {\tilde N}_0\ = \ N_0 \ +\  6\log\left[ \frac {3\Gamma(1/3)^6}{16\pi ^4} \right]\ .
	\ee
	Here $N_0$ is the number of species at the 'desert point', which essentially gives us the minimal possible number 
	of species in the tower governed by the modulus ${\rm Im}\tau=\tau_2$. This number is model dependent but one would expect it in the range of a few $\sim 100$. In this example, it depends on the K3 geometry.
	One thus has for the species scale
	 \cite{vandeHeisteeg:2023dlw,Castellano:2023aum}
	\beq 
	\Lambda^2 \ \sim \  \frac {M_{\rm P}^2}{N} \ \simeq \ \frac {M_{\rm P}^2}{-6 \log(\tau_2 |\eta(\tau)|^4) \ +\ {\tilde N}_0}\ .
	\eeq
	Upon going to a canonical kinetic term for $\tau$, this is a positive function, symmetric with respect to the $\tau_2=1$ axis and exponentially falling on both sides, see figure \ref{fig:speciesvspotential}. This is an interesting example which we will
 use as a proxy below.

 Given a string theory vacuum with a number of moduli parametrized by $\phi_i$ one can define the  species
 scale gradient quantity $\lambda(\phi_i)$ as \cite{Calderon-Infante:2023ler}
 \be
\lambda(\phi_i) \ =\ G^{ij}   \frac { ( \partial_i \Lambda)(\partial_j \Lambda) } { \Lambda^2 }  \ ,
\ee
with $G^{ij}$ the (inverse) metric in moduli space. This quantity measures the rate of variation of the
species scale along the different directions in moduli space and it was conjectured it is bounded from below 
by \cite{Calderon-Infante:2020dhm,Calderon-Infante:2023ler}.
\beq
\lambda (\phi)  \ \geq \ \frac {1}{\sqrt{(d-1)(d-2)}}
\label{constante1}
\eeq
with $d$ the number of dimensions. This lower bound is saturated when we take 'decompactification limits' in moduli space,
in which some extra dimension(s) open. In directions corresponding to 'emergent string' limits it is given by the stronger
bound \cite{vandeHeisteeg:2022btw,vandeHeisteeg:2023ubh,Calderon-Infante:2023ler}
\beq
\lambda (\phi)  \ \geq \ \frac {1}{\sqrt{d-2}},
\label{constante2} 
\eeq
and is saturated. Thus e.g. in the case of the Enriques CY $K3\times T^2/{\bf Z}_2$, the limit with the 
torus volume going to infinity ${\rm Im}\tau\rightarrow \infty$ is an emergent string limit. The string arises upon a NS5-brane wrapping the $K3$ surface.

  The functions that have been found as coefficients of higher dimensional BPS operators are
  non-holomorphic Eisenstein series automorphic forms
  associated to the duality symmetries in the theory \cite{Green:2010kv,Green:2010wi,Green:1999pv,Pioline:1998mn,Green:1999pu,Green:2005ba}. One important property 
  of these Eisenstein objects is that they have so called 'constant terms', i.e. polynomial terms in their expansion
  \cite{Green:2010kv,Green:2010wi}. This
  property underlies the bounds eqs.(\ref{constante1},\ref{constante2}) above. In the following, we will find that our inflaton potentials will be related to non-holomorphic Eisenstein forms, and the existence of these ‘constant terms’ will be the root of the existence of an inflaton ‘plateau’.

	\section{A proposal for inflaton potentials}\label{sec:proposal}
	
	Our aim in this paper is to show that the fact that the fundamental scale of the theory is  $\Lambda$ (rather than $M_{\rm P}$), 
	that it is moduli dependent and also an automorphic form,  can guide us in order to propose inflaton
	potentials that are consistent with Quantum Gravity constraints. 
	
	In identifying consistent potentials we will impose the following conditions:
	
	\begin{itemize}
	
	\item 1) V is an invariant automorphic form of the duality symmetries of the theory.
	
	\item  2) The potential should  have no singularities at a finite distance
	in moduli space.
	
	\item 3) The potential depends only on $\Lambda$ and its field derivatives.
	
	\item 4) We will impose that at least in some large moduli direction the potential reaches a plateau.
	
	\end{itemize}

	The condition 2) is imposed so that the EFT makes sense.  Simplicity suggests condition 3)  but also the fact that for
	e.g., with a single modulus, the species scale contains all the information relevant to the tower arising at a large modulus.
	Finally, condition 4) is imposed so that in the asymptotics there is a flat potential that could give rise to a
	sufficiently large period of inflation. Due to the properties of the species scale discussed in the previous section,
 and in particular to condition 4),
a natural proposal is as follows:  
\beq
\boxed {  V(\phi_i) \ =\ \mu^{d-2}M_{\rm P}^2 \lambda(\phi_i)\ =\   \mu^{d-2}M_{\rm P}^2\,
G^{ij} \,  \frac { ( \partial_i \Lambda)(\partial_j \Lambda) } { \Lambda^2 } } \ .
\label{gen}
\eeq
 Thus we are promoting the generator of the convex hull of the species scale to a scalar potential.
 
 As we said, it was conjectured and checked in string theory vacua that this quantity $\lambda$ verifies the bounds
 \cite{Calderon-Infante:2023ler} above 
and in fact the bound are saturated asymptotically in different examples. So in this way condition 4) is fulfilled.
Also the species scale $\Lambda$ is an (invariant)  automorphic form and so is $\lambda$ (condition 1)).
It only depends on $\Lambda$ and its first derivatives so that all the conditions above are attained.	

Our motivation up to here has been bottom-up, independent of any specific String theory vacua,
nor the presence or not of supersymmetry.
On the other hand, the above potential is positive definite thus, 
e.g. in the context of ${\cal N}=1,2$ $d= 4$ SUSY  would only arise from a 
D-term potential. Such a D-term potential may be motivated if we assume e.g. that the modulus is in a vector multiplet
and the  K\"ahler potential for the
moduli in the vector multiplet behaves like $K\sim -\log N$. 
We will see momentarily that e.g.
in the specific CY examples the asymptotic behaviour of $K$ is correct since 
(e.g. for single modulus) $N\rightarrow \tau_2$.
Then a D-term potential would arise from
\be
V_D \ \simeq \ D^2 \ \simeq \  G^{\tau{\bar \tau}}\frac {\partial K}{\partial\tau}
\frac {\partial K}{\partial{\bar \tau}}\ \simeq \ 
 G^{\tau{\bar \tau}} \frac {1}{N^2} \frac {\partial N}{\partial\tau}
\frac {\partial N}{\partial{\bar \tau}}\ \simeq \ 
 G^{\tau{\bar \tau}}\frac {1}{\Lambda^2}\frac {\partial \Lambda}{\partial\tau}
\frac {\partial \Lambda}{\partial{\bar \tau}}\ \ ,
\ee
 which indeed corresponds to the potential in equation (\ref{gen}) for $d=4$.
 It is interesting to remark that D-term potentials in the context of 
 supergravity leading to potentials similar to Starobinsky were constructed
 in \cite{Farakos:2013cqa,Ferrara:2013rsa}. The potentials appear as D-terms associated to
 a massive vector multiplet in the context of 'new minimal supergravity'. 
 Thus loosly speaking, our potential would look like a modular invariant generalization 
 of that class of supergravity models with a massive vector multiplet.
In any event, in this paper  we will remain agnostic about the 
  origin of such a potential and consider it as an explicit example of a function
 which 1) takes into account that the fundamental scale $\Lambda$ is field dependent,
 2) Modular invariant (so takes dualities into account) and 3) gives rise to a successful model
 of inflation (due to the plateau at large moduli). 

Let us note that such potentials cannot be valid over the whole moduli space, even if we restrict ourselves to the
fundamental region of the relevant duality symmetries. In particular, there is a natural bound
that one expects to be obeyed, 
\be
V(\phi) \ \lesssim \Lambda^d \ ,
\label{cota}
\ee
if one wants to remain within an EFT approach. So we will impose this condition when
considering the inflaton potential.

\section{ A modular invariant inflaton potential}\label{example}
	
	A specific realistic string vacuum will have a 'universal species scale'  $\Lambda$ depending on all moduli. The idea is that 
	at some point in the evolution of the universe all but one complex modulus $\tau$ were fixed in their minima. 
	This modulus could have different physical interpretations e.g. K\"ahler modulus, complex structure modulus
	, or complex dilaton in a  Calabi-Yau (CY) compactification. Since the mathematics is the same, we will not make any distinction at this level.
	Let us consider then  the case of a single complex modulus $\tau=\tau_1 + i \tau_2$ with an action
		\be
	{\cal L}_{mod} \ =\   \frac {2}{a^2}M_{\rm P}^2\frac {|\partial_\mu  \tau|^2}{(\tau-\tau^*)^2} \ \ -\ V(\tau,\tau^*) \ .
	\ee
	The kinetic term normalisation $a$  depends on the type of infinite distance limit  $\tau\rightarrow \infty $ that we are taking
 see e.g.\cite{vandeHeisteeg:2023ubh}.
	In particular when $\tau $ corresponds to e.g. the overall volume of a compactification one has $a=\sqrt{2/3}$.
	The simplest and best studied modular symmetry for a complex modulus is $SL(2,{\bf Z})$, which arises in 
	$T^2$ string compactifications (or orbifolds thereof) as well as elliptic fibrations. 
	We will thus assume $\tau$  to be a modulus transforming under $SL(2,{\bf Z})$ in the usual way
 (see the Appendix \ref{app:modular} for some useful formulae concerning the relevant modular forms).
	The  kinetic term above  is modular invariant and  the corresponding canonically normalised real scalar $\phi$ is related to $\tau_2$ by
	\beq
	{\rm Im}\tau\ = \  \tau_2 \  =\ e^{a\phi/M_{\rm P}} \ .
 \label{canonico}
	\eeq
 As an explicit example of species scale in a 4d setting, we will take the case mentioned above corresponding to 
  Type IIA compactification in  Enriques CY manifolds of the form $K3\times T^2/{\bf Z}_2$. Very similar results 
  would be obtained for other CY manifolds which e.g. could however present invariance under other $SL(2,{\bf R})$ discrete 
  subgroups. Also some 4d ${\cal N}=1$ CY compactifications obtained by some modding/projection of those ${\cal N}=2$
  compactifications could feature similar structures. Thus, as we said,  in this example we have
\be
	N \ \simeq \ -6 \log\left[2\tau_2|\eta(\tau)|^4\right] \ +\  {\tilde N}_0 \  \ ;\ \Lambda^2 \ \simeq \ \frac {M_{\rm P}^2}{N} \ .
 \ee
A natural proposal for the scalar potential is thus
\begin{equation}
   V =  \mu^2 M_{\rm P}^2 \ G^{\tau{\bar \tau}}\frac{\partial_\tau \Lambda }{\Lambda}
 \frac {  \partial_{\bar \tau} \Lambda}{\Lambda} \ \simeq \ 
 \mu^2 M_{\rm P}^2 \ G^{\tau{\bar \tau}}\frac{\partial_\tau N }{N}
 \frac {  \partial_{\bar \tau} N}{N} \ ,
 \label{eq: potential}
\end{equation}
where $\mu$ is a mass parameter which will eventually be fixed by the amplitude of scalar perturbations.
The derivative of the number of species $N$ with respect to the holomorphic variable $\tau$ takes the  form: 
\begin{equation}
    \frac{\partial N}{\partial \tau} = -12 \frac{\eta^{'}(\tau)}{\eta(\tau)} - \frac{6}{\tau - \Bar{\tau}} = \frac{3}{\pi i}\left(\text{\textbf{G}}_2(\tau)- \frac{\pi}{ \tau_2}\right) \equiv \frac{3}{\pi i}\tilde{\text{\textbf{G}}_2}(\tau,\bar{\tau})
\end{equation}
where we have used $\tilde{\text{\textbf{G}}_2} \equiv \text{\textbf{G}}_2 - \pi/\tau_2$ and $\text{\textbf{G}}_2 =- 4 \pi i \,\log(\eta)/\partial \tau$, see Appendix \ref{app:modular} for more details.
Therefore, the derivative with respect to $\tau_2$ reads
\begin{equation}
    \frac{\partial N}{\partial \tau_2} = \frac{6}{\pi} {\rm Re}\,\tilde{\text{\textbf{G}}}_2(\tau) \ .\label{eq:derN}
\end{equation}
Moreover, inserting this into \eqref{eq: potential} and taking into account $G^{\tau\Bar{\tau}}=a^2(\tau-\bar{\tau})^2/2$, we arrive at \footnote{From now on, we refer to $G_n$ as the real part of the Eisenstein's series \textbf{G}$_n$}
\begin{equation}
  \boxed { V = \frac{9a^2 \mu^2M_{\rm P}^2}{\pi^2 } \left(\frac{\tau_2\,\tilde{G}_2}{N}\right)^2} \label{potentialnd}
\end{equation}
It is interesting to see the behaviour at large modulus $\tau_2$. One has
\be
G_2\ \longrightarrow \ \frac {\pi^2}{3} \ ;\ \frac {N}{\tau_2} \ \longrightarrow \ 2\pi ,
\ee
so that
\beq
V\ =\ \mu^2M_{\rm P}^2\ \lambda(\tau,\tau^*) \ \longrightarrow \ \frac {a^2}{4}\ \mu^2 M_{\rm P}^2 \ .
\ee
Note that, as expected, the limit is a constant, which depends on the normalisation of the
modulus. For $a= \sqrt{2/3}$ one gets a factor $1/6$ whereas for the case $a=\sqrt{2}$  one gets a
factor 1/2. The former corresponds to a decompactification limit whereas in the other case, it corresponds to 
an emergent string limit see \cite{vandeHeisteeg:2023ubh,vandeHeisteeg:2023dlw}. If we keep the next term in the $1/\tau_2$ expansion one gets 
\be
V\ \longrightarrow \ \left(\frac {3a}{2\pi}\right)^2 \mu^2 M_{\rm P}^2 |\frac {\pi}{3}\ -\ e^{-a\phi}|^2 \ .
\ee
Interestingly, for large (real) modulus the potential takes a form very similar to that of the 
Starobinsky model for $a=\sqrt{2/3}$, see also the attractor models
of \cite{Kallosh:2013yoa,Kallosh:2015zsa},
so it will not be very surprising that we get very similar results 
for the inflaton parameters.
The correspondence is not identical (although $\pi/3\simeq 1$) since for small moduli the potentials start to
behave differently. Furthermore, the modular invariant potential will be symmetric around $\phi=0$ 
(see figure \ref{fig:modularvstaro}) whereas the Starobinsky one presents a wall.

An interesting point to make is that in a modular invariant theory like this, the physics will  be restricted to
a fundamental region of $SL(2,{\bf Z})$. Thus $\tau_2 \in [\sqrt{3}/2,\infty)$ (although we will also get an upper bound for $\tau_2$
from physical considerations, see below). However the axionic field range is compact $\tau_1 \in [-1/2, 1/2]$. 
Due to this, in the region in which inflation is produced $\tau_2\gg \tau_1$ and one can ignore the effects of $\tau_1$.
Thus in the computation of inflation parameters we will set $\tau_1=0$.

\begin{figure}[h!]
    \centering
    \includegraphics[width=0.7\linewidth]{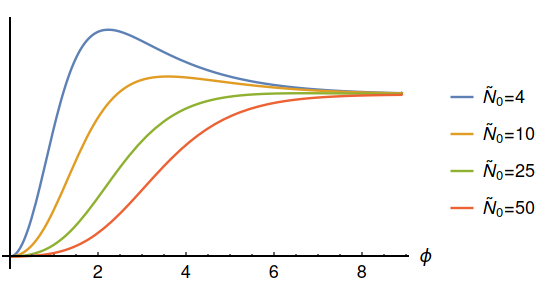}
    \caption{The different values of the scalar potential of \eqref{potentialnd} as a function of the number of species at the desert point.
    Only the right side of the potential with $\phi> 0$ is shown. We use $a=\sqrt{2/3}$ for these plots.}
    \label{fig:potentialNd}
\end{figure}

At small $ \tau_2$ the value of $N_0$, the number of species at the 'desert point' we mentioned above,  
starts also to play a role. In fact, along the real line direction with $\tau_1=0$ that we will consider here
to describe inflation, the potential presents maxima close to the $\tau_2\sim 2$ region in which the potential deviates 
from a 'Starobinsky behaviour'. This is illustrated in figure (\ref{fig:potentialNd}) in which the scalar potential
$V(\tau_2)$ is shown for different values of the 'desert point' number ${\tilde N}_0$. One finds that the 
position of this maximum is pushed to the right as ${\tilde N}_0$ increases so that for values 
${\tilde N}_0\gtrsim 25$ the maxima disappear and the potential becomes monotonously decreasing to the left, in the 
region in which the EFT is consistent. The existence of this maxima for the convex hull generator $\lambda$ in 
Type IIA CY compactifications was already noticed \cite{vandeHeisteeg:2023ubh,vandeHeisteeg:2023dlw}. There it was also noticed that in general 
the presence of a ${\tilde N}_0>0$ is in fact required, and typically may depend on other moduli ($K3$ moduli in the case at hand).
Its actual value will depend on the compactification and we thus will take it as a free parameter. 
In following computations we will take for definiteness ${\tilde N}_0= 25$. However, the inflaton parameters are 
practically independent of the value as long as ${\tilde N}_0\lesssim 100$.

\section{Inflation parameters}\label{sec:inflationpara}

Let us now compute the inflation parameters of the above potential. 
As argued above,  to compute the inflaton parameters we will 
 set $\tau_1=0$ in the explicit expressions. 

\subsection{ The \texorpdfstring{$\epsilon$}{Lg} parameter}

Let us rewrite the potential as 
\begin{equation}
    V= K^2 \left(\frac{\tau_2 \tilde{G}_2}{N}\right)^2 \ ;\ K\ \equiv \ \frac {3a}{\pi}\ \mu M_{\rm P}
    \label{potseminal}
\end{equation}
Its derivative takes the following form 
\begin{equation}
    \frac{\partial V}{\partial \tau_2} = 2K^2 \left(\frac{\tau_2 \tilde{G}_2}{N}\right) \left[\frac{\tilde{G}_2}{N}+ \frac{\tau_2}{N}\frac{\partial \tilde{G}_2}{\partial \tau_2} - \frac{\tau_2 \tilde{G}_2}{N^2}\frac{\partial N}{\partial \tau_2}\right].
\end{equation}
Now, for the second term we know that 
\begin{equation}
    \frac{\partial \tilde{G}_2}{\partial \tau_2} = \frac{\partial G_2}{\partial \tau_2} + \frac{\pi}{\tau_2^2}.
\end{equation}
Using a  modular form relationship, see Appendix \ref{app:modular}, we obtain
\begin{equation}
    \frac{\partial \text{\textbf{G}}_2}{\partial \tau} = \frac{1}{2 \pi i}\left(5  \text{\textbf{G}}_4 -  \text{\textbf{G}}_2^2\right)
\end{equation}
and from there
\begin{equation}
    \frac{\partial \tilde{G}_2}{\partial \tau_2} = \frac{1}{2 \pi }\left(5 G_4 - G_2^2\right) + \frac{\pi}{\tau_2^2}.
\end{equation}
For the third term we use \eqref{eq:derN}. Putting it all together we arrive at the following expression
\begin{equation}
  \frac{\partial V}{\partial \tau_2} = 2K^2 \left(\frac{\tau_2 \tilde{G}_2}{N}\right) \left[\frac{G_2}{N}+ \frac{\tau_2}{2 \pi N}\left(5 G_4 - G_2^2\right)  - \frac{6\tau_2 \tilde{G}_2^2}{\pi N^2}\right].
\end{equation}
Going now to a canonical kinetic term as in eq.(\ref{canonico}) 
and substituting $K$ one gets
\be
     \frac{\partial V}{\partial \phi} \ = \ 
     \frac{6 a^2 } {\pi} \mu M_{\rm P}\,\tau_2 \sqrt{V} \left[\frac{G_2}{N}+ \frac{\tau_2}{2 \pi N}\left(5 G_4 - G_2^2\right) - \frac{6\tau_2 \tilde{G}_2^2}{\pi N^2}\right] \notag
\ee
From these expressions, we can compute  the slow roll parameter $\epsilon$, defined as 
\begin{equation}
    \epsilon = \frac{1}{2}\left(\frac{V^{'}}{V}\right)^2,
\end{equation}
and thus,

\be
\boxed  {
\epsilon\ =\ \frac {2a^2 }{({\tilde G}_2)^2}\left[G_2 \ +\ \frac {\tau_2}{2\pi} (5G_4-G_2^2)\ -\ \frac {6\tau_2}{\pi N} 
{\tilde G}_2^2\right]^2 } \ .
\ee
This is an interesting formula that gives us an explicit modular invariant expression for $\epsilon$ in the full
range of the inflaton values under a consistent  EFT action.  One can check numerically it has the
the shape of a bell with a maximum at the origin $\phi=0$ and exponentially decreasing on both sides. 
For large $\phi$ the above expression 
greatly simplifies. There is a cancellation 
\begin{equation}
    G_2 - \frac{6 \tau_2 G_2^2}{\pi N} \hspace{1em}\xrightarrow{\tau_2>>1}\hspace{0.5 em} 0 \ ,
\end{equation}
where we have used 
\begin{equation}
    G_2 \rightarrow \frac{\pi^2}{3},\quad \frac{N}{\tau_2}\rightarrow 2\pi \ .
\end{equation}
The derivative of the potential with respect to $\phi$, in the large $\phi$ limit reads
\begin{equation}
     \frac{\partial V}{\partial \phi} \simeq \frac{a K^2 \pi}{3}\left(-\frac{6 \tau_2^2}{\pi N^2}\left[\frac{\pi^2}{\tau_2^2} - \frac{2\pi G_2}{\tau_2}\right]\right)\simeq  \frac{2 a}{3}\frac{K^2 \pi^2}{N}.
\end{equation}
From here, the slow roll parameter $\epsilon$ reads 
\begin{equation}
    \epsilon \simeq 2\frac{ (12 a)^2}{N^2} \ .
\end{equation}
Let us now compute the number of e-folds before the end of inflation.
This can be computed from the formula
\begin{equation}
    N_e =  \int_{\phi_{f}}^{\phi_i}\, \frac{V}{V^{'}} {\rm d}\phi = \frac{1}{\sqrt{2}}\int_{\phi_{f}}^{\phi_i}\, \frac{1}{\sqrt{\epsilon}}{\rm d}\phi.
\end{equation}
Obtaining an analytical expression becomes cumbersome, so we will content ourselves here with 
the results in the limit $N\rightarrow 2\pi \tau_2$ which is enough for our purposes since most of inflation 
takes place in this regime\footnote{A numerical calculation without any approximation yields the same results.}. One gets
\begin{equation}
    N_e = \frac{1}{24 a} \int_{\phi_{f}}^{\phi_i}\, N {\rm d}\phi = \frac{\pi}{12 a^2} \left(\tau_2^i - \tau_2^f\right).
\end{equation}
Since one has  $\tau_2^i\gg \tau_2^f$ one has 
\begin{equation}
    \tau_2^i \simeq \frac{12a^2 N_e}{\pi} \ ,
\end{equation}
and for the canonical field 
\be
\phi_i \ \simeq \ \frac {1}{a} \log(12a^2N_e/\pi) \ .
\ee
For $N_e= 60$ one has $\phi_i\simeq 6.1(4.3)$ for $a=\sqrt{2/3}(a=\sqrt{2})$.

Note that for large $\tau_2$ there is an interesting relationship between the number of e-folds $N_e$ and 
the number of species $N$. One has
\be
\boxed {
    N_e \simeq \frac{N}{24\, a^2} }.
\ee
This is interesting since it relates the number of e-folds produced to the number of species $N$ in the tower.
This seems to suggest that the entropy stored in the tower containing 
$N$ species is spent in the creation of $N_e$ e-folds of volume. 
Note also that for $N_e\simeq 60$, one has $N\simeq 1440$, so that  $N\gg {\tilde N}_0$ and the 
species at the desert point have no effect on the computation of the inflation parameters.

Note also that in terms of $N_e$, using the above expression, $\epsilon$ has a simplified asymptotic value:
\be
\epsilon \ \simeq \ \frac {1}{2a^2N_e^2} \ .
\ee

\begin{figure}[h!]
    \centering
\includegraphics[width=0.6\linewidth]{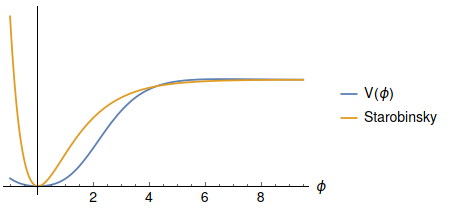}
    \caption{The shape of the Starobinsky potential (orange) and the modular invariant potential at $\tau_1=0$ (blue).}
    \label{fig:modularvstaro}
\end{figure}

\subsection{Normalization}

To obtain the normalization of the potential, i.e. the value of $\mu$, one can make use of the experimentally measured value for the
amplitude of scalar perturbations ${\cal A}_s$. The amplitude may be written in terms of the potential data at $N_e$ e-folds before the end of 
inflation. One has
\be
{\cal A}_s \ \simeq \ \frac {1}{24\pi ^2}\frac {V}{\epsilon M_{\rm P}^4} \ .
\ee
At large $\tau_2$ one obtains
\be
{\cal A}_s \ \simeq \ \frac {a^4N_e^2}{48\pi^2} \frac {\mu^2}{M_{\rm P}^2} \ ,
\ee
and then  for the scale $\mu$ 
\be
\frac {\mu }{M_{\rm P}} \ \simeq \ {\cal A}_s^{1/2} \frac {4\sqrt{3} \pi}{a^2N_e} \ .
\ee
Experimentally, ${\cal A}_s\simeq 2.2\times 10^{-9}$ \cite{Planck:2018jri} so that one gets for $N_e=60$,
\be
\mu \ \simeq \ M_{\rm P} \frac {0.36}{a^2} {\cal A}_s^{1/2} \ \simeq \ \frac {10^{13.5} } {2a^2} GeV \ .
\ee
As we said above,
we must require to have the potential below the species scale so that an EFT makes sense.
In particular, we must have 
\begin{equation}
    V\ \lesssim \  \Lambda ^4
\end{equation}
For large $\tau_2$ this means \begin{equation}
    K^2 \frac{\tau_2^2\,\tilde{G}_2^2}{N^2} \ \lesssim \ \frac{1}{N^2} \ \longrightarrow \
    \tau_2 \ \lesssim \ \frac{1}{K \tilde{G}_2} \ \longrightarrow \ \frac{M_{\rm P}}{\pi a \mu}  \ .
\end{equation}
Numerically one finds the canonically normalized 
scalar field $\phi$ 
\be
|\phi|  \ \lesssim \ \frac {1}{a} \log(\frac {2a}{\pi}\ 10^{4.5})\ \simeq \ \frac {10}{a}.
\ee
Thus e.g. for $a=\sqrt{2/3}$ ($a=\sqrt{2}$) one gets $|\phi|\lesssim 12 (7)$. These values are about twice as large as the 
initial values for $\phi_i$ to get the sufficient inflation we found above.  Thus this condition substantially
constraint the initial conditions in the inflation process.

\subsection{ The \texorpdfstring{$\eta$}{Lg} parameter and the spectral index}

To compute $\eta$ we need to take the second derivative of the potential in the canonical frame.
This becomes quite involved, so let us compute it first in the large inflaton limit
(for the exact expression see eq.(\ref{masainfla})).
At leading order, for large $\tau_2$ the second derivative of the potential $V$ takes the following form
\begin{equation}
    \frac{\partial^2 V}{\partial \phi^2} \simeq \frac{\partial }{\partial \phi}\frac{2 a K^2 \pi^2}{3 N} = -\frac{2 a K^2 \pi^2}{3 N^2} \frac{\partial N}{\partial \phi} = -\frac{2 a K^2 \pi^2}{3 N^2} \frac{6 \tilde{G}_2}{\pi}\frac{\partial \tau_2}{\partial \phi} \simeq -\frac{2 a^2 K^2 \pi^2}{3 N}
\end{equation}
Therefore, 
\begin{equation}
    \eta_V = \eta + \epsilon = \frac{V^{''}}{V} = - \frac{24 a^2}{N}.
\end{equation}
We can then compute the  spectral index $n_s$ as 
\begin{equation}
    n_s =1 -6 \epsilon_V + 2 \eta_V \simeq 1- \frac{ (12)^3 a^2}{N^2} -\frac{48 a^2}{N}  \ .
\end{equation}
Writing it in terms of the e-fold number through $N=24 a^2N_e$ one gets
\be
n_s \simeq 1\ -\ \frac {3}{a^2N_e^2}\ -\ \frac {2}{N_e} \ \simeq \ 1\ -\ \frac {2}{N_e}  \ .
\ee
Note that this quantity is independent of $a$.
For  $N_e = 60(50)$ one then gets
\begin{equation}
    n_s = 0.966 (0.960)
\end{equation}
in remarkable agreement with experiment \cite{Planck:2018jri}.
The tensor to scalar ratio $r=16\epsilon$ takes then the value 
\begin{equation}
    r = \frac{4608 a^2}{N^2} \ \simeq \ \frac {8}{a^2N_e^2} =0.0033(0.0011)
\end{equation}
for $N_e=60$ and $a=\sqrt{2/3},\,\sqrt{2}$ respectively. This may be within reach of future CMB observations.

\subsection{Summary of inflation parameters}

We have seen that {\it all inflation parameters may be written in terms of the number of species $N$ associated with
the tower of states becoming light when the inflaton is large}. Alternatively, given that $\Lambda^2\simeq M_{\rm P}^2/N$, 
all parameters may be written in terms of the species scale evaluated at the beginning of inflation.
Thus one has:
\be
\epsilon \ \simeq \ \frac {2(12a)^2}{N^2}\  \ ;\ \eta \ \simeq \ -\frac {24 a^2}{N} ;\ r\  \simeq \ \frac {4608a^2}{N^2}
\ee
\be
{\cal A_s} \simeq \ \frac {\mu^2}  {72\pi^2M_{\rm P}^2}\ N^2 \ ;\ n_s\   \simeq \ 1 - \frac {48a^2}{N}  - \frac {(12)^3a^2}{N^2}
\ee
On the other hand, the number of species $N$ determines the number of produced e-folds through $N=24a^2N_e$. Writing these results
in terms of $N_e$ simplify a bit and one gets
\be
\epsilon \ \simeq \ \frac {1}{2a^2N_e^2}\  \ ;\ \eta \ \simeq \ -\frac {1}{N_e} ;\ r\  \simeq \ \frac {12}{N_e^2} \ ; 
n_s\   \simeq \ 1 - \frac {2}{N_e}  - \frac {3}{a^2N_e^2} \ .
\ee
Therefore, using $a=\sqrt{2/3}$ one precisely recovers the results for Starobinsky inflation, see  Appendix A.

\vspace{1em}

\subsection{Mass of the inflaton at the minimum}

In order to compute the value of $\eta$ at inflation it is sufficient to take the $\tau_2\rightarrow \infty $
limit as we did above. 
However, the second derivative of the potential may be of interest to compute the inflaton mass's value at the end of inflation.
We give here for completeness the (a bit involved) exact expression
\bea
   \frac{\partial^2 V}{\partial \phi^2} =& \frac{a^2 K^2}{2 \pi^2 N^4}\Big[\left(\pi^2 + 5 \tau_2^2 G_4\right)^2 N^2 + 70 \tau_2^4 \tilde{G}_2 G_6 N^2 + 3 \tau_2^4 \tilde{G}_2^4 \left(144+20 N+N^2\right)\\
   &-4 \tau_2^2 \tilde{G}_2^2 N \left(5 \tau_2^2 G_4(2 N+15)+\pi ^2 (N+15)\right)\Big],\notag
   \label{masainfla}
\eea
where we have used, see Appendix \ref{app:modular} for more details,
\be
\frac{\partial \text{\textbf{G}}_4}{\partial \tau} = \frac{1}{i\pi}\left(7 \text{\textbf{G}}_6 - 2 \text{\textbf{G}}_2 \text{\textbf{G}}_4\right).
\ee
The value of the second derivative at $\phi=0$ takes the form
\begin{equation}
    \frac{\partial^2 V}{\partial \phi^2}|_{\phi=0} = \frac{a^2 K^2}{2 \pi^2 N^2}\left(\pi^2 + 5 G_4\right)^2|_{\phi=0}
\end{equation}
where we have used that $G_6=\Tilde{G}_2=0$ at $\phi=0$. Moreover, we know that
\begin{equation}
    N|_{\phi=0}-N_0 = -6 \log 2 - 24 \log\left(\frac{\Gamma(\frac{1}{4})}{2 \pi^{3/4}}\right)\simeq 2.16925,\quad G_4|_{\phi=0}\simeq 3.15121.
\end{equation}
Therefore, 
\begin{equation}
     \frac{\partial^2 V}{\partial \phi^2}|_{\phi=0} = \frac{30.3363\,a^4 \mu^2}{ (2.16925 + N_0)^2} \ ,
\end{equation}
so that e.g. for $a=\sqrt{2/3}$ and $N_0=25$ we arrive to
\begin{equation}
    \frac{\partial^2 V}{\partial \phi^2}|_{\phi=0} = 0.01826 \,\mu^2.
\end{equation}
The Strarobisnky potential predicts a mass$^2$ for the inflation of $V^{''}=\mu^2$. 
Here we find a mass factor $7.4$ lighter,  given that our Starobinsky-like modular function is less steep at the origin.  This is due to the fact that their potentials differ when approaching the origin. In particular, the contribution from $N_0$ is no longer negligible and its effect is to flatten the potential 
around the origin. This is observed in fig.(\ref{fig:modularvstaro}) in which the modular invariant potential with $a=\sqrt{2/3}$ 
is plotted along with the Starobinsky potential. Most of inflation is produced in the flat region $\phi_i=6\rightarrow 4$,
in which the potentials behave identically. Even in the region $\phi_i=3\rightarrow 1$ the potentials are 
relatively parallel so the $\epsilon $ in both cases becomes similar.

\subsection{Some comments on inflation dynamics}

There are a number of interesting points to remark concerning inflation in this 
setting.

\begin{itemize}

\item In standard inflation the entropy at the beginning of inflation is very small, and entropy increases till the end of 
inflation. In our case the initial entropy is huge because having a large inflaton vev like $\phi_i\sim 5-6 M_{\rm P}$ implies there is
a tower of states (e.g. KK or winding) with $N$ species becoming light. On the other hand as $\phi$ decreases, also the number of
species $N$ decreases and so does entropy. Since the total entropy cannot decrease, this means that some presumably thermal
entropy is created as $\phi$ decreases. In fact we have seen that the number of e-folds is proportional to $N$, $N_e\sim N$.
Then by the end of inflation, the volume has grown like ${\rm Vol}\sim e^{3N_e}\sim e^{3N/2\pi a^2}$, with the volume determined
by the initial amount of entropy stored in the tower. It would be interesting to better understand this connection.

\item 
In standard plateau potentials like Starobinsky's the range of it extends in principle to infinity,
and, as usual,  there is a fine-tuning for the initial state. 
In our case the condition bounding the potential like $V\lesssim \Lambda^4$ for the EFT to make sense,
forces the inflaton to be in a narrow range $\phi \in [-10/a , 10/a ]M_{\rm P}$. The value $\phi_i < (5\sim 6)M_{\rm P}$ required
to get at least $50-60$ e-folds is comfortably within this range. However the range of possible initial conditions 
is now much restricted. In particular, close to the upper boundary of $\tau$ the kinetic energy should be vanishing small so that
the total energy does not exceed the $\Lambda$ cut-off. 

\item 
One of the criticisms for the Starobinsky inflation is that in principle there could be important 
corrections when the inflaton becomes trans-Planckian. Thus e.g. in principle one should include higher powers of the 
curvature to the action like ${\cal R}^{2n}, n>1$. When this is done, the nice properties of the Starobinsky potential, like flatness,
 are in general spoiled.  In our case we have not derived the potential from such type of
Lagrangians. The idea is that the effective potential 
includes already all relevant corrections, like is the case of the species scale functions computed in \cite{vandeHeisteeg:2022btw,vandeHeisteeg:2023ubh,Castellano:2023aum,vandeHeisteeg:2023dlw}.
Furthermore, the property of modular invariance makes it difficult to envisage possible 
sensible (e.g. non-singular) corrections.

\item 
There are a number of points that deserve further study. One of them is the origin of the $\mu$ parameter which sets the 
the overall scale of the potential, which we fix to be of order $\mu\simeq 10^{14}$GeV in order to accommodate CMB data. 
This is not constrained by modular invariance and possibly would be determined by the dynamics of the other
moduli in the theory. It would also be interesting to study the possible role of $\tau_1$ as $\tau_2$ becames small.
This would be particularly important around the two types of minima appearing in this potential at $\tau=(\pm 1 + i \sqrt{3})/2$.
Finally, we have not addressed so far reheating, which would require some knowledge of the couplings of the modulus $\tau$ to SM physics which will be very much model dependent.

\end{itemize}

Let us also comment at this point that 
 there have been previous studies of modular invariant effective potentials in the context of ${\cal N}=1$ 
 supergravity since a long time, see  e.g. \cite{ferrara:1989bc,Font:1990nt,Font:1990gx,Nilles:1990jv,Cvetic:1991qm}
 and  \cite{Leedom:2022zdm,Ding:2024neh,Qu:2024rns,Abe:2023ylh,Lynker:2019joa,King:2024ssx,Ding:2020zxw,
 Kobayashi:2016mzg,Ben-Dayan:2008fhy,Covi:2008cn,Gaillard:1998xx}. 
 There is also more recent work related to Swampland conditions \cite{Gonzalo:2018guu,Cribiori:2023sch}.
 All  these in general address the generation 
 of potentials from F-term contributions.
 However, it seems difficult to get a potential of the form we consider in this note  starting with 
 F-terms in a ${\cal N}=1$ action, since our potential is positive definite. The sugra vacua obtained from F-terms
 typically (although not always) live in AdS and show no asymptotic plateau.
Thus e.g. the modular invariant examples obtained for gaugino condensation in the context of
heterotic compactifications tend  to grow exponentially at infinity with the (un-normalized) complex modulus
\cite{Font:1990nt,Cvetic:1991qm,Gonzalo:2018guu}. In general the regime of validity of these and other 
modular invariant examples should be restricted then as in eq.(\ref{cota}).

\section{ Final comments and outlook}\label{conclusions}

Many phenomenologically successful models of inflation involve trans-Planckian excursions of the inflaton.
However, it has become clear in recent years that taking such large values for fields implies the presence of
towers of states becoming light,  which can ruin the consistency of the EFT. So taking such large values for scalars has to be done
with some care. In addition, it has also been realized that the notion of fundamental scale 
(at which Quantum Gravity cannot be ignored) is moduli dependent. The fundamental 'Species Scale' $\Lambda $
of a QG vacuum is in general an automorphic form of the duality symmetries of the theory. 

Guided by these ideas,  in this paper we have looked for candidate inflaton potentials in which the inflaton is
a modulus in some  QG/String Theory compactification.  We were interested in inflatons with trans-Planckian 
excursions,  so we know there will be some associated tower of states becoming light. The properties of such a tower are
encapsulated in its contribution to the Species Scale $\Lambda$.  We thus search for a potential which
1) Is a modular invariant function of the duality symmetries of the theory, 2) Depends only on the 
Species Scale $\Lambda$ and its derivatives, which carry the information about the tower and 3) 
At large moduli has a plateau behaviour, so that  sufficient inflation is produced.  This last property is very constraining.
We argue that the simplest functions of the moduli  with the required properties (particularly the 3rd) are 
$V\sim \lambda (\phi_i , \phi_i ^*)$,
with $\lambda = G^{i{\bar j}}  (\partial_i \Lambda )(\partial_{\bar j} \Lambda) /\Lambda^2$, the generator
of the 'convex hull' of the Species Scale gradients. In particular it has been shown recently that 
$\lambda(\phi_i,\phi_i^*) \longrightarrow const,$ when moduli become large, which is what we need
to get a plateau. 
Mathematically this behaviour arises because the Species Scale is given by 
Eisenstein Series forms which feature 'constant (polynomial) terms'  in their expansion \cite{Green:2010kv,Green:2010wi}.
In some sense the idea in practice would be to
promote the generator $\lambda$ to a scalar potential, at least in what concerns the inflaton potential.

We propose then to identify the inflaton in 4d with the imaginary part $\tau_2$ 
of a complex scalar modulus $\tau$  (like e.g. the K\"ahler, complex structure or
complex dilaton in CY String compactifications) with a modular invariant scalar potential. To get an idea of 
the structure of $\Lambda$ for  the tower asociated to the inflaton in a realistic case,  we take as a 
proxy a Type IIA compactification  on an Enriques CY of the form $K3\times T^2/{\bf Z}_2$, with a torus volume modulus
$\tau$ transforming under a $SL(2,{\bf Z})$ duality symmetry.  Taking the potential 
proportional to the convex hull generator $\lambda(\tau,\tau^*)$ for this case one finds the remarkable simple
expression $V=K^2|\tau_2{\tilde G}_2|^2\Lambda^4$. Surprisingly, when one takes the large inflaton ($\tau_2$) limit
one obtains a scalar potential very similar to the Starobinsky  potential. This unexpected result is interesting, because
we know that the Starobinsky results for inflaton parameters are in remarkable agreement with experiment. However,
for general values of $\tau$ the inflaton parameters have specific modular invariant expressions depending on
Eisenstein series forms, and differ from that of Starobinsky.
In the large inflaton limit, one also obtains modular invariant expressions for $\epsilon , \eta$
and $n_s$. Thus one obtains $\epsilon\simeq \Lambda^4$ and $\eta\simeq -\Lambda^2$ and the remarkable result that
the number of species in the tower at the beginning of inflation ($N$) is proportional to the number of e-folds 
obtained, $N\simeq 2\pi a^2N_e$, with $a$ the normalization of the modulus. This suggests inflaton dynamics in which 
at the beginning of inflation there is a large entropy stored in the tower of states, and as $\phi$ decreases, this entropy drops and presumably should be overtaken by some form of thermal entropy.  

The potential  eq.(\ref{potseminal}) we propose is quite unique and leads to very successful 
inflaton predictions. Its correspondence with the generator of the convex hull Species Scale gradient is very intriguing,
although we do not have an explanation of why such a generator should be promoted to a scalar inflaton potential. On the other hand,
as we explain above, it could perhaps be understood as a D-term potential. It would be interesting to
check whether such a potential admits such a  D-term interpretation. In the meantime, 
we can consider the potential as a simple inflaton potential example with characteristics compatible with what we
know of quantum gravity, like modular invariance, the role of towers,  the species scale, and entropy.

 Given the nice relationship between inflation parameters and entropy-related quantities, like the number of species $N$
 and the species scale $\Lambda$, it is tempting to speculate that such a relation may be actually more general than the
 particular derivation we have worked out here. Thus, e.g. even in a situation with multiple active moduli and
 towers, the species scale is a single unique function, so one could speculate that the expressions
 \be
 N_e \ \simeq \  1/(24a^2\Lambda^2) \ ;\  \epsilon\ \simeq \ 2(12a)^2\Lambda^4 \ ;\ \eta \ \simeq \ -24a^2\Lambda^2 \ ;\ n_s \ \simeq \ 1 \ -\ 48a^2\Lambda^2
 \ee
 could have more general validity. They would correspond to Starobinsly-like numerical predictions, but 
 more general and with intriguing connections
 with entropy-related quantities.
 \\
\\
\\
\\

\centerline{\bf \large Acknowledgements}

\vspace*{.5cm}

We thank Jose J. Blanco-Pillado, Alberto Castellano, \'Alvaro Herr\' aez, Fernando Marchesano and  \'Angel Uranga for discussions.  This work is supported through the grants CEX2020-001007-S and PID2021-123017NB-I00, funded by MCIN/AEI/10.13039 /501100011033 and by ERDF A way of making Europe. G.F.C. is supported by the grant PRE2021-097279 funded by MCIN/AEI/ 10.13039/501100011033 and by ESF+. 


                 
\appendix

\section{Starobinsky inflation}
\label{onedimension}

The Starobinsky model\cite{Starobinsky:1980te} of inflation  (see e.g. \cite{DiMarco:2024yzn,Mohanty:2020pfa,Baumann:2014nda}) starts from a non-minimal gravity action of the form
\be
S \ =\ \ \frac {M_{\rm P}^2}{2}\int d^4x \sqrt{-g} \left(R \ +\ \frac {1}{6M^2} R^2\right)
\label{Staraction3}
\ee
After a Weyl rescaling and field redefinitions this action turns out to be equivalent to Einstein's gravity coupled to a single
real scalar $\phi$ and a scalar potential of the form
\be
V_{Staro} \ =\ \frac {3}{4} M^2M_{\rm P}^2\left(1\ - \ e^{a\phi/M_{\rm P}}\right)^2 \ ,
\ee
with $a=\sqrt{2/3}$.
Applying the equations of motion the scalar curvature may be expressed in terms of this scalar as
\be
R \ =\ 6 M^2{ N}_{Star} \ ;\ {N}_{Star}(\phi) \ \equiv \ (e^{a\phi}\ - \ 1) .
\ee
The form of the potential is shown in fig. \ref{fig:modularvstaro} in the main text.  From the potential one can easily compute the inflaton parameters and one gets (see e.g. \cite{Baumann:2014nda})
\be
\epsilon_V\ =\ \frac { 2a^2}{N_{Star}^2 } \ ;\ 
\eta_V \ =\ -\frac {2a^2(N_{Star}-1)}{N_{Star}^2} \ ; \ n_s \ \simeq \ 1-\frac {4a^2(N_{Star}+2)}{N_{Star^2}}\ ;\ r\ =\ \frac {32a^2}{N_{Star}^2}
\ee
For the number of e-folds one obtains
\be
N_e(\phi) \simeq \frac{1}{2a^2} \left( N_{Star}(\phi) - N_{Star}(\phi_{end}) - \frac{1}{2a}(\phi - \phi_{end}) \right).
\ee
At large $\phi$ one has $N_e\simeq N_{Star}/(2a^2)$, so we can write the parameters in terms of $N_e$ as
\be
\epsilon_V\ =\ \frac { 1}{2a^2N_{e}^2 } \ ;\ 
\eta_V \ =\ -\frac {1} {N_e}  \ ; \ n_s \ \simeq \ 1-\frac {2}{N_e}\ ;\ r\ =\ \frac {8}{a^2N_{e}^2} \ ,
\ee
with $a=\sqrt{2/3}$.

Since the large field behaviour of the Starobinsky potential and that of the modular invariant potential is similar,
the results for the inflaton parameters are the same. Note also that we have here defined the function $N_{Star}$
in analogy with the number of species $N$ in the modular invariant model. At the large field limit the correspondence would be
$N=12N_{Star}$. 

There is also some analogy between eq.(\ref{desarrollo} and eq.(\ref{Staraction}), in that both 
involve ${\cal R}^2$ actions. An important difference is that
in the Starobinsky action the coefficient $M$ is a fixed, modulus-independent scale. On the contrary, 
one expects in quantum gravity that the coefficient of the ${\cal R}^2$ term should be a field dependent
function. Another crucial difference is that in the Starobinsky model the inflaton is a degree of freedom
induced by the higher curvature term. In our case the inflaton is an explicit modulus in the theory.
Furthermore the ${\cal R}^2$ action in eq.(\ref{desarrollo}) is a just tool to compute what is the 
field dependent Species Scale $\Lambda$ in the theory.

\section{Modular forms for \texorpdfstring{${\rm SL}(2,\mathbb{Z})$}{Lg}}\label{app:modular}

The modular group ${\rm SL}(2,\mathbb{Z})$ lies discretely in 
${\rm SL}(2,\mathbb{R})$. It consists of integer $2\times 2$ matrices with unit determinant, and its action on the upper half plane $H$, takes the following form
\be
{\rm SL}(2,\mathbb{Z}) \times H \rightarrow H, \quad (\Gamma, \tau) \equiv \frac{a\tau + b}{c \tau + d}, \quad {\rm where}\,\, \Gamma=\begin{pmatrix}
a & b \\
c & d
\end{pmatrix}.
\ee
The generators of the group are given by the two matrices 
\be
T=\begin{pmatrix}
1 & 1 \\
0 & 1
\end{pmatrix},\quad S=\begin{pmatrix}
0 & -1 \\
1 & 0
\end{pmatrix},
\ee
whose action on $\tau$ is given by $\tau\rightarrow \tau +1$ and  $\tau\rightarrow -\frac{1}{\tau}$ respectively.  

A modular form of weight $k\in \mathbb{Z}$ on $\Gamma$ is a function $f:H\rightarrow \mathbb{C}$ with a Fourier expansion $f(\tau)=\sum_{n=0}^{\infty}a(n)e^{2\pi i \tau}$ which converges for all $\tau\in H$ and satisfies the following modular transformation properties \cite{apostol,Zagier2008}:
\begin{itemize}
    \item For any $\Gamma\in {\rm SL},(2,\mathbb{Z})$, we have \begin{equation}
        f(\Gamma\cdot\tau) = (c\tau + d)^k\,f(\tau)\ .
    \end{equation}
    \item For any $\Gamma\in {\rm SL},(2,\mathbb{Z})$ the function $f(\tau)$ satisfies \begin{equation}
        |f(\tau)|<\infty, \quad {\rm as}\,\, {\rm Im}\tau \rightarrow \infty \ . 
    \end{equation}
\end{itemize}

A clear example of modular form is the \textbf{Eisenstein series}. These are defined as \cite{Zagier2008}
\be
\text{\textbf{G}}_k = \sum_{(m,n)\in\mathbb{Z}^2} \frac{1}{(m\tau + n)^k},\quad {\rm with}\,\, (m,n)\neq (0,0).
\ee
It is a modular form of weight $k$ which vanishes identically for $k$ odd and converges absolutely for $k>2$. They can also be rewritten in the following form,
\be
\text{\textbf{G}}_k = 2\zeta(k) + 2\sum_{m=1}^{\infty}\sum_{n\in \mathbb{Z}}\frac{1}{(m\tau + n)^k},
\ee
such that for ${\rm Im}\tau\rightarrow\infty$, $\text{\textbf{G}}_k\rightarrow 2\zeta(k)$.\newline 
The case $k=2$ is special due to its non-convergence and therefore does not represent a pure modular form. It is usually denoted as \textit{quasimodular} and has the following transformation property,
\be
\text{\textbf{G}}_2(\Gamma\cdot\tau) = (c\tau + d)^2\text{\textbf{G}}_2(\tau)-\pi i c(c\tau+d).
\ee
One might wonder whether a complete version of the Eisenstein series $k=2$ still exists. By taking the limit $G_{2(1+\epsilon)}$ with $\epsilon\rightarrow 0$, it can be shown that the non-holomorphic function 
\be
\tilde{\text{\textbf{G}}}_2 \equiv \text{\textbf{G}}_2 - \frac{\pi}{{\rm Im}\tau},
\ee
does the job \cite{Zagier2008}. This is the so-called \textit{almost holomorphic modular form} of weight 2. 
The case $k=2$ can also be expressed in terms of the $\eta(\tau)$ function as
\be
\text{\textbf{G}}_2 = -4\pi i \frac{{\rm d}}{{\rm d}\tau} \log \eta(\eta).
\ee

An interesting property of these Eisenstein series is that their derivatives can be expressed in terms of higher and lower Eisenstein series. Here we write a few examples \cite{Cvetic:1991qm,Zagier2008}
\bea
\frac{{\rm d}}{{\rm d}\tau}\text{\textbf{G}}_2 &=& \frac{1}{2\pi i}\left(5\text{\textbf{G}}_4 - \text{\textbf{G}}_2^2\right),\\
\frac{{\rm d}}{{\rm d}\tau}\text{\textbf{G}}_4 &=& \frac{1}{i\pi}\left(7\text{\textbf{G}}_6-2\text{\textbf{G}}_2\text{\textbf{G}}_4
\right),\\
\frac{{\rm d}}{{\rm d}\tau}\text{\textbf{G}}_6 &=& \frac{1}{i \pi}\left(10\text{\textbf{G}}_8-3\text{\textbf{G}}_2\text{\textbf{G}}_6\right).
\eea
Finally, we write some series expansions we use along the text. Given $q=e^{2\pi i \tau}$,
\bea
\eta(\tau)&=& q^{1/24}\left(1-q -q^2 + q^5 + q^7-q^{12}-\dots\right)\\
\text{\textbf{G}}_2  &=& 2\zeta(2)\left(1-24 q - 72 q^2 - 96q^3 - 168 q^4 - \dots\right)\\
\text{\textbf{G}}_4  &=& 2\zeta(4)\left( 1 + 240 q + 2160 q^2 + 6720 q^3 + 17520 q^4 + \dots\right)\\
\text{\textbf{G}}_6  &=& 2\zeta(6)\left(1-504 q - 16632 q^2 - 122976 q^3 - 532728 q^4-\dots\right)
\eea

\bibliographystyle{JHEP2015}
\bibliography{bibliography}

\end{document}